\documentclass[prd,letterpaper,twocolumn,preprintnumbers,nofootinbib,superscriptaddress]{revtex4}

\usepackage{amsmath,amssymb}
\usepackage{graphicx}
\usepackage{units}
\usepackage{bbold}
\usepackage{xcolor}
\usepackage{dsfont}
\usepackage[hyperfootnotes=false,colorlinks,citecolor=blue]{hyperref}

\usepackage{comment}
\usepackage[normalem]{ulem}     

\newcommand{\dd}{{\rm d}}
\newcommand{\ii}{{\rm i}}

\def \epsilon {\varepsilon}

\newcommand{\orcid}[1]{\href{https://orcid.org/#1}{#1}}

\begin{document}

\title{Scalar baryons in neutron stars}

\author{Julian Heeck}
\email[E-mail: ]{heeck@virginia.edu}
\thanks{\orcid{0000-0003-2653-5962}}
\affiliation{Department of Physics, University of Virginia,
Charlottesville, Virginia 22904, USA}

\author{Yu Zhi}
\email[E-mail: ]{yz9hy@virginia.edu}
\thanks{\orcid{0009-0007-3440-4678}}
\affiliation{Department of Physics, University of Virginia,
Charlottesville, Virginia 22904, USA}

\begin{abstract}
Neutron stars have baryon chemical potentials that can exceed the neutron mass, providing favorable conditions to convert neutrons to new particles carrying baryon number, even if those processes are kinematically forbidden in vacuum. We study the effect of GeV-scale scalars with baryon number on neutron stars' equation of state and show that non-perturbatively large repulsive self-couplings are required to support the observed two-solar-mass neutron stars. We also study potentials with attractive self-interactions, which can trigger scalar production even for masses above the chemical potential and resemble Coleman's Q-matter.
\end{abstract}

\maketitle

\section{Introduction}
\label{sec:intro}

Particles beyond the Standard Model that carry baryon number can have interesting phenomenology, notably as (asymmetric) dark matter candidates or in connection to the matter--antimatter asymmetry of our universe. Simple models can be found in Refs.~\cite{Heeck:2020nbq,Strumia:2021ybk,Heeck:2025uwh}, the most popular being a new neutral fermion $\chi$ with baryon number $B(\chi)=1$ and effective interactions such as the dimension-six portal $\bar{\chi}udd/\Lambda^2$, which leads to a mass mixing of $\chi$ with the neutron $n\sim udd$ at low energies. For $m_\chi < m_n$, neutrons can then decay into this lighter neutral baryon, e.g.~via $n\to\chi\gamma$, which is subject to strong constraints~\cite{Davoudiasl:2014gfa,Fornal:2018eol} that become even stronger for lighter $\chi$ once the channels $n\to \chi \pi^0$ and $p\to \chi \pi^+$ open up~\cite{Super-Kamiokande:2013rwg,Heeck:2025uwh}. Aside from the phase-space suppressed region $m_\chi\simeq m_n$~\cite{Fornal:2018eol,Pfutzner:2018ieu,McKeen:2020zni,Fornal:2023wji} -- where \textit{bound} neutrons are stable -- neutron decay rates into $\chi$ are suppressed to at least $\unit[10^{-30}]{yr^{-1}}$ via invisible-neutron searches~\cite{Heeck:2019kgr,SNO:2022trz}, to be further improved by JUNO~\cite{JUNO:2024pur}.

For $m_\chi \geq m_n$, nucleons are stable and the model is much less constrained, but as shown in Refs.~\cite{McKeen:2018xwc,Baym:2018ljz}, $\chi$ could still be produced in neutron stars as long as $m_\chi$ is smaller than the \textit{chemical potential} $\mu_B$ of neutrons/baryons, which can be of order $\unit[1.5]{GeV}$ depending on the Equation of State (EoS). The conversion $n\to\chi$ into a non-interacting $\chi$ then relieves Fermi pressure, drastically softening the neutron star's EoS and typically lowering the maximal mass a neutron star can have. Our observations of old neutron stars with masses around  $2 M_\odot$, e.g.~PSR J1614-2230~\cite{Demorest:2010bx} and~PSR J0348+0432~\cite{Antoniadis:2013pzd}, then put constraints on this new-physics model. Similar analyses have been performed in Refs.~\cite{Cline:2018ami,Grinstein:2018ptl,McKeen:2020oyr,Motta:2018bil,Gardner:2023wyl}, usually in the context of the neutron-lifetime anomaly~\cite{Wietfeldt:2011suo,SciAm2016}, which requires a sizable free-neutron conversion branching ratio $\text{BR}(n\to \chi)\simeq 1\%$.

In this letter, we will apply a similar analysis to \emph{scalar} baryons, i.e.~complex scalar fields $\phi$ with $B(\phi)=1$~\cite{McKeen:2020zni,Heeck:2020nbq,Khatibi:2023fwv,Heeck:2025uwh}. The simplest low-energy couplings are of the form $n \nu \phi^*$ and $p e \phi^*$, which induce $n\to \phi \bar{\nu}$ or $p\to \phi e^+$ if the new scalar is light, again constrained to rates below  $\unit[10^{-30}]{yr^{-1}}$~\cite{Heeck:2025uwh}. For $m_\phi \geq m_n$, the conversion rates and underlying couplings could be large enough to affect neutron stars, as long as $m_\phi < \mu_B$. Once again, the neutron's Fermi pressure can be relieved by transforming a neutron into the new particle, but now the new particle will not fill up its own Fermi sphere but just condense in the core since it follows Bose statistics. The $\phi$ would not generate any pressure in this case and soften the neutron star's EoS even more dramatically than the fermion $\chi$! However, scalars generically have repulsive self-interaction terms to stabilize the scalar potential, e.g.~$\lambda |\phi|^4$, so $\phi$ will \emph{not} end up in a point-like core after all but instead form a sphere due to the self-interaction pressure, rendering the EoS stiffer. Just like in the fermion case, the observation of  $2 M_\odot$ neutron stars then puts constraints on the $m_\phi$--$\lambda$ parameter space.

Our setup differs from the often discussed scenario of dark-matter-admixed neutron stars, which are usually treated in a two-fluid approximation~\cite{Karkevandi:2024vov,Grippa:2024ach}. In our case, new particles are not \textit{captured} by neutron stars but \textit{produced} inside, and hence insensitive to the neutron star's environment. Since we assume chemical equilibrium of all baryons, the equations are also much simpler. Lastly, the particles under discussion are not dark matter, as they can decay into nucleons in the region of interest.

\section{Neutron Stars}

The interior of neutron stars is described by the Tolman--Oppenheimer--Volkoff (TOV) equation~\cite{Tolman:1939jz,Oppenheimer:1939ne},
\begin{align}
    \frac{\mathrm{d}P}{\mathrm{d}r} &=  -\frac{m}{r^2}\epsilon \left(1 + \frac{P}{\epsilon } \right) \left(1+\frac{4\pi r^3 P}{m}\right)\left(1-\frac{2m}{r}\right)^{-1} ,\label{eq:tov}
\end{align}
where $r$ is the radial coordinate measured from the center of the star, $\epsilon (r)$ is the energy density, and $P(r)$  the pressure; we work in natural units where $c = \hbar =G= 1$. The gravitational mass, $m$, is defined as the mass measured by the gravitational field felt by a distant observer, and is determined by the (energy) density through the continuity equation:
\begin{align}
    \frac{\mathrm{d}m}{\mathrm{d}r} & =4\pi r^2 \epsilon\,.
\end{align}
Once the EoS of the neutron-star matter is specified by some relation $\epsilon (P)$, the equations can be solved subject to the boundary conditions $m(0)=0$ and $P(0)=P_c$, where $P_c$ is the core pressure. The neutron-star radius $R$ is defined as the distance where the pressure drops to zero, i.e.~$P(R) = 0$, which also gives the star's mass as $M = m(R)$. 

By varying the central pressure $P_c$ for a given EoS, one can generate mass--radius curves $M(R)$ that can be compared to neutron-star observations, and, in particular, feature a \emph{maximal} neutron star mass~\cite{Oppenheimer:1939ne}, which must be compatible with the observed $2 M_\odot$ neutron stars~\cite{Demorest:2010bx,Antoniadis:2013pzd}. In this way, we obtain constraints on the underlying EoS, which are notoriously difficult to calculate even in the absence of new physics because they involve nuclear physics in an extremely dense environment.
Here, we employ three different EoS from the database CompOSE~\cite{CompOSE:website, Typel:2013rza, Oertel:2016bki, CompOSECoreTeam:2022ddl}, following  common practice in modern numerical relativity, e.g.~\cite{Servignat:2023lai, Camilletti:2022jms}:
\begin{itemize}
    \item \textbf{APR} unified crust~\cite{Akmal:1998cf, Davis:2024nda, Haensel:2007yy, Davis:2025nwz}, named after Akmal, Pandharipande, and Ravenhall~\cite{Akmal:1998cf}, which is a microscopic EoS obtained using variational many-body methods with the Argonne ($v_{18}$) two-nucleon interaction, relativistic boost corrections, and the Urbana IX three-nucleon interaction. We use the CompOSE version in which the neutron-star crust is reconstructed in a thermodynamically consistent manner.
    \item \textbf{HS(DD2)}~\cite{Hempel:2009mc, Typel:2009sy, Moller:1996uf, Hempel:2011mk, Steiner:2012rk}, which combines the nuclear statistical-equilibrium model of Hempel and Schaffner-Bielich with the density-dependent relativistic mean-field interaction DD2.
    \item \textbf{RG(SLy4)}~\cite{Danielewicz:2008cm, Gulminelli:2015csa, Chabanat:1997un}, which denotes the Raduta–Gulminelli~\cite{Gulminelli:2015csa} unified EoS constructed using the non-relativistic SLy4 (``Skyrme Lyon 4'') energy-density functional.
\end{itemize}
For a given EoS, the TOV equation~\eqref{eq:tov} can be solved numerically, although this requires some care given the stiffness of the differential equation. Following the approach of Ref.~\cite{Lindblom:1998dp}, we reformulate the equations by defining the enthalpy variable
\begin{equation}
    h(P) = \int_0^P \frac{\dd P'}{\epsilon (P') + P'}\,,
\end{equation}
which simply tracks the baryon chemical potential $\mu_B$ since  $h = \ln(\mu_B/m_n)$.
Defining further $u(h) = r(h)^2$ and $v(h) = m(h)/r(h)$, the enthalpy-based TOV equations take the form
\begin{align}
    \frac{\mathrm{d}u}{\mathrm{d}h} &= -\frac{2u(1-2v)}{4\pi uP(h) + v}\,, \\
    \frac{\mathrm{d}v}{\mathrm{d}h} &= -\frac{(1-2v)(4\pi u \epsilon(h) - v)}{4\pi uP(h) + v}\,.
\end{align}
At the center, $h = h_c$, $u(h_c)$ = 0 and $v(h_c) = 0$; at the surface, $h = 0$, $u(0) = R^2$, and $v(0) = M/R$. The equations are still singular at the center, so we use the analytical expansion from Ref.~\cite{Lindblom:1998dp},
\begin{align}
    u(h) &= \frac{3(h_c - h)}{2\pi (\epsilon_c + 3 P_c)} + \mathcal{O}\left[(h_c - h)^2\right],\\
    v(h) &= \frac{2\epsilon_c (h_c - h)}{\epsilon_c + 3P_c} + \mathcal{O}\left[(h_c - h)^2\right],
\end{align}
and start the numerical integration at a tiny non-zero radius. 
The above allows us to generate mass--radius curves for each EoS, which agree with the curves provided by CompOSE within our desired accuracy.

\section{Neutron Star with Dark Fermion}

Before delving into the scalar baryon case, let us review the familiar case of a new fermionic baryon $\chi$~\cite{McKeen:2018xwc,Baym:2018ljz}, often dubbed \textit{dark fermion}. We assume chemical equilibrium $\mu_B = \mu_\chi$, i.e.~sufficiently fast conversion rates $n\leftrightarrow \chi$ compared to the lifetime of the neutron star, which we take to be of order of a billion years.
Neutrons can then relieve pressure due to the Pauli exclusion principle and strong nuclear interactions by converting into the non-interacting $\chi$, which fills up its own Fermi sphere up to Fermi momentum $k_\chi$, with energy density $\epsilon_\chi$ and pressure $P_\chi$ given by
\begin{align}
    k_\chi & = \sqrt{\mu_B^2 - m^2_\chi}\,,\\
    \epsilon_\chi & = \frac{1}{\pi^2}\int_0^{k_{\chi}} \mathrm{d}k\, k^2 \sqrt{k^2 + m_\chi^2}\,,\\
    P_\chi &= -\epsilon_\chi + \mu_Bn_\chi\,,
\end{align}
with number density $n_\chi = k_\chi^3/(3\pi^2)$~\cite{McKeen:2018xwc,Baym:2018ljz}. 
Importantly, a $\chi$ population requires $\mu_B > m_\chi$, and since $\mu_B > m_n$ inside the dense neutron-star core, $\chi$ particles can be produced even though the vacuum decay $n\to \chi$ would be kinematically forbidden.
For $\mu_B\gg m_\chi$, i.e.~for large pressure, the $\chi$ EoS is approximately
\begin{align}
    \epsilon_\chi \simeq 3 P_\chi\,,
    \label{eq:FermionEoS_largeP}
\end{align}
whereas in the low-pressure regime, or non-relativistic limit, $\mu_B\simeq m_\chi$, we have
\begin{align}
    \epsilon_\chi \simeq \frac{5^{3/5}}{3^{2/5}\pi^{4/5}} m_\chi^{4} \left(\frac{P_\chi}{m_\chi^4}\right)^{3/5} ,
    \label{eq:FermionEoS_smallP}
\end{align}
and adding up these two right-hand side expressions gives a surprisingly good approximation for $\epsilon_\chi (P_\chi)$ over the entire pressure range, off by at most $5\%$; we nevertheless use the full numerical approach in what follows.

The total EoS consisting of standard nuclear matter and the new fermion $\chi$ is then given by 
\begin{align}
    P_\text{tot} & = P_\text{nuc} + P_\chi \,,\\
    \epsilon_\text{tot} & = \epsilon_\text{nuc} + \epsilon_\chi\,.
\end{align}
CompOSE provides $\mu_B (n_B)$, $P_\text{nuc}(n_B)$, $\epsilon_\text{nuc}(n_B)$ for a given EoS. We have $n_\chi$, $P_\chi$, and $\epsilon_\chi$ as functions of $\mu_B$, so this gives us $P_\text{tot}$, $\epsilon_\text{tot}$, and $n_\text{tot}$ as a function of the nuclear baryon density $n_B$, i.e.~the parametric EoS $P_\text{tot}(\epsilon_\text{tot})$ required to solve the TOV equation.

\begin{figure}[tb]
    \centering
    \includegraphics[width=0.94\linewidth]{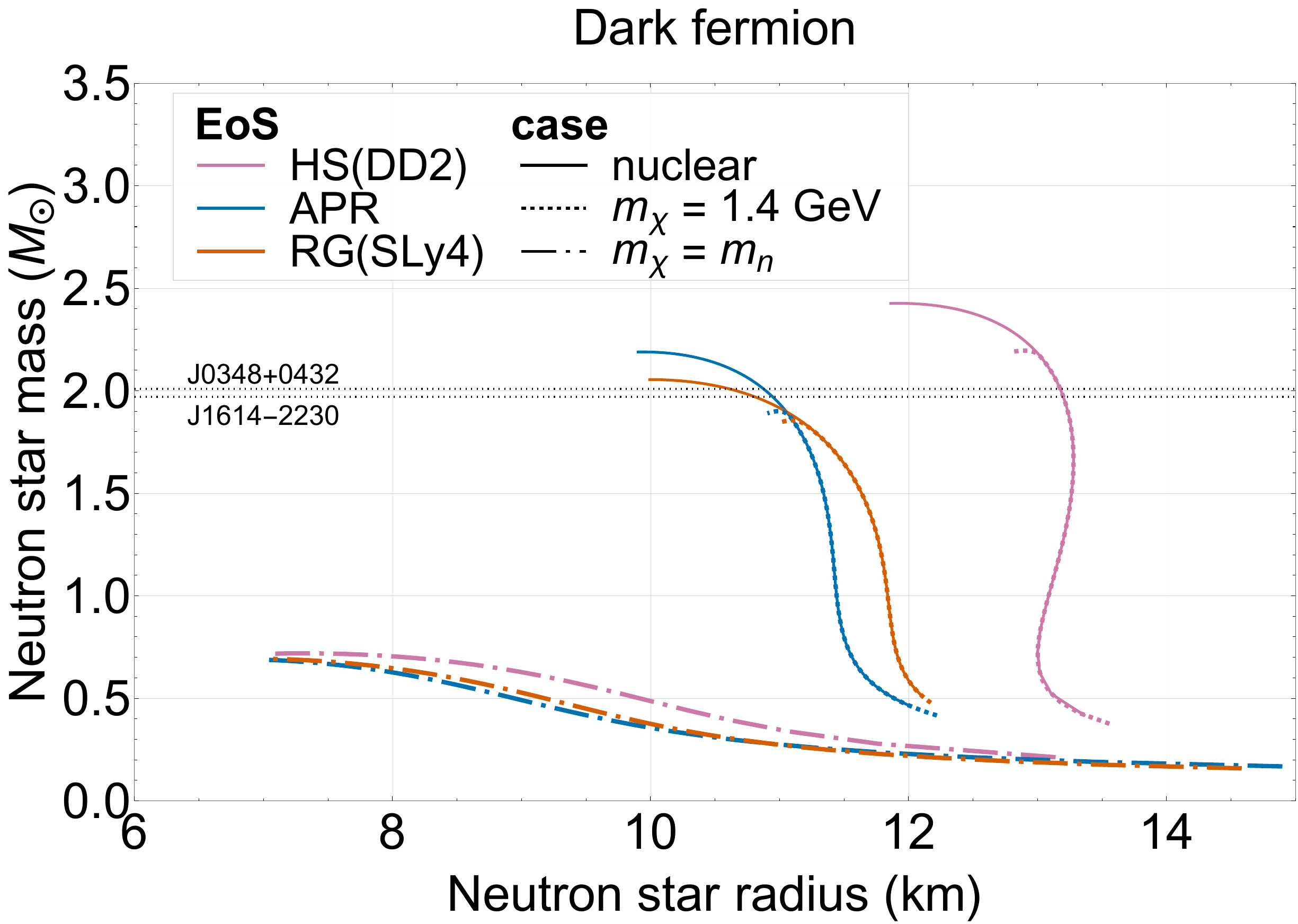}
    \caption{Neutron star with a dark fermion $\chi$ for three different nuclear EoS (colors) and three different $\chi$ masses (dashes and dots). Curves are clipped such that they end at their maximum mass.}
    \label{fig:dark_fermion}
\end{figure}

We show the resulting mass--radius curves in Fig.~\ref{fig:dark_fermion}, in qualitative agreement with Refs.~\cite{McKeen:2018xwc,Baym:2018ljz}. Without $\chi$, or for $m_\chi\to \infty$, all three of our chosen EoS allow for neutron-star masses above $2 M_\odot$ (solid lines), but the addition of $\chi$ can drastically lower the maximal mass: for $m_\chi \leq m_n$ neutron stars would not even reach one solar mass (dot-dashed lines), confidently excluded by observations. Pushing $m_\chi$ above $m_n$ restricts the $n\to \chi$ transition to the neutron-star core, where $\mu_B > m_n$, alleviating these constraints somewhat (dotted lines). For HS, APR, and RG EoS, fermion masses below $1.3$, $1.5$, and $\unit[1.7]{GeV}$ are excluded, respectively. Circumventing these limits requires either additional repulsive self-interactions of~$\chi$~\cite{McKeen:2018xwc,Baym:2018ljz,Cline:2018ami,Grinstein:2018ptl,Motta:2018bil} or transition rates $n\leftrightarrow \chi$ far below the neutron star's age, say below $\unit[10^{-9}]{yr^{-1}}$.

\section{Neutron Star with Dark Boson}

While even a non-interacting dark fermion still contributes Fermi pressure to the neutron star's EoS, a non-interacting dark boson $\phi$ would cause a runaway decay $n\to\phi$ into a pressure-less $\phi$ condensate, trivially excluding the model unless $m_\phi > \mu_B^\text{max}$ or the transition rates  are too slow. The situation is more interesting than that, though, since scalars in quantum field theory are generically allowed self-interaction terms in the scalar potential, e.g.~the typical repulsive $\lambda |\phi|^4$ interaction, which then generates pressure and allows for a stable equilibrium inside the neutron star. For large-enough $\lambda$, the transitions $n\to \phi$ will even be blocked, although we will see below that this requires large coupling constants far outside the perturbative regime.

Following the derivation from Ref.~\cite{Suarez:2016eez}, we model the dark bosons as a zero-temperature complex scalar condensate with the standard Lagrangian density
\begin{equation}
    \mathcal{L}_\phi = |\partial \phi|^2 -m_\phi^2|\phi|^2 - V_\text{int}(|\phi|)
\end{equation}
with some interaction potential $V_\text{int}(|\phi|)$ that ensures conservation of the assumed $U(1)$ baryon symmetry of $\phi$, i.e.~that the full potential has a global minimum at $\phi=0$. Not shown are the terms that connect $\phi$ and the nucleons, e.g.~$n \nu \phi^*$ or $p e \phi^*$, which we assume to be sufficiently efficient to induce $n\to \phi$ transitions over the lifetime of the neutron star and equilibrate baryons. 
In the local-density or Thomas--Fermi approximation, spatial gradients of the condensate amplitude are neglected. A stationary condensate may then be written as 
\begin{equation}
    \phi(t) = \sqrt{X}e^{-\ii\mu_\phi t}\,,
\end{equation}
where $\mu_\phi$ is the boson chemical potential. The Lagrange field equation gives
\begin{equation}
    \mu_\phi^2 = m^2_\phi + \frac{\mathrm{d}V_\text{int}}{\mathrm{d}X}\,,
    \label{eq:EOM}
\end{equation}
which determines the condensate amplitude $X$ as a function of $\mu_\phi$. The global $U(1)$ symmetry of the complex scalar field gives a conserved current, and its temporal component gives the conserved $\phi$ number density 
\begin{equation}
    n_\phi = 2\mu_\phi X\,.
\end{equation}
For a homogeneous condensate, the energy density and pressure obtained from the energy-momentum tensor are~\cite{Suarez:2016eez}
\begin{align}
    \epsilon_\phi &= (\mu_\phi^2 + m_\phi^2)X + V_\text{int}(X)\,,\\
    P_\phi &=  (\mu_\phi^2 - m_\phi^2)X - V_\text{int}(X)\,.
\end{align}
which satisfy the same zero-temperature thermodynamic relation as the dark fermion:
\begin{equation}
    \epsilon_\phi + P_\phi = \mu_\phi n_\phi\,.
\end{equation}
In the next subsections we will study a few concrete examples for scalar self-interactions and their impact on neutron stars.
Notice that all these models also support pure-$\phi$ boson stars, balancing the gravitational attraction with the repulsive self-interaction, see Refs.~\cite{Colpi:1986ye,Chavanis:2011cz}.

\subsection{Monomial interactions}

Considering a simple repulsive monomial self-interaction term $V_\text{int} = \lambda_N|\phi|^N/m_\phi^{N-4}$ with $N>2$ and positive dimensionless coefficient $\lambda_N$, the field equation~\eqref{eq:EOM} is solved for the  condensate amplitude
\begin{equation}
    X(\mu_\phi) = m_\phi^2 \left(\frac{\mu_\phi^2 - m_\phi^2}{N \lambda_N m_\phi^2/2}\right)^{\frac{2}{N-2}} ,
\end{equation}
leading to
\begin{align}
    P_\phi &= \frac{N-2}{N}(\mu_\phi^2 - m_\phi^2)X(\mu_\phi) \,,\\
    \epsilon_\phi &= \frac{(N+2)\mu_\phi^2 + (N-2)m^2_\phi}{N}X(\mu_\phi)\,.
\end{align}
The relativistic sound speed $c_{s,\phi}$ is then given by
\begin{equation}
    c^2_{s,\phi} = \frac{\partial P_\phi /\partial X}{\partial \epsilon_\phi / \partial X} = \frac{(N-2)(\mu_\phi^2 - m_\phi^2)}{(N+2)\mu_\phi^2 - (N-2)m^2_\phi}\,,
\end{equation}
which goes to $0$  for $\mu_\phi \to m_\phi$ and $(N-2)/(N+2)$ for $\mu_\phi \to \infty$, and is thus always subluminal.
The $\phi$ EoS takes the simple form
\begin{align}
    \epsilon_\phi = \frac{N+2}{N-2}P_\phi+ 2 m_\phi^4 \left(\frac{2 P_\phi}{(N-2)\lambda_N m_\phi^4}\right)^{2/N},
\end{align}
which matches the dark-fermion EoS in the large pressure regime~\eqref{eq:FermionEoS_largeP} for $N=4$, but the small-pressure EoS is inevitably different from the fermionic case~\eqref{eq:FermionEoS_smallP} if we restrict ourselves to integer $N$.\footnote{Otherwise, the low-pressure EoS of dark fermion and scalar would match for $N=10/3\simeq 3.33$ and $\lambda= \sqrt[3]{972\pi^4}/5\simeq 9.12$.}
In the non-relativistic limit, $\epsilon_\phi \simeq m_\phi n_\phi$, and the pressure is parametrically suppressed:
\begin{align}
    \frac{P_\phi}{\epsilon_\phi} \simeq \frac{N-2}{2^{(N+2)/2}}\,\lambda_N \left(\frac{n_\phi}{m_\phi^3}\right)^{(N-2)/2} .
\end{align}
For GeV-scale $m_\phi$ and $n_\phi$ of order of the neutron density, $n_\phi \simeq (\unit[100]{MeV})^3$, one needs $\lambda_N\gg 1$ for a stiff EoS, i.e.~$P_\phi\sim \epsilon_\phi$, growing exponentially with $N$: $\lambda_3 \sim 200$, $\lambda_4\sim 4000$, $\lambda_6\sim 4\times 10^6$. Our numerical analysis below confirms the need for such large couplings to support $2 M_\odot$ neutron stars.

\begin{figure*}[tb]
    \centering
    \includegraphics[width=0.48\linewidth]{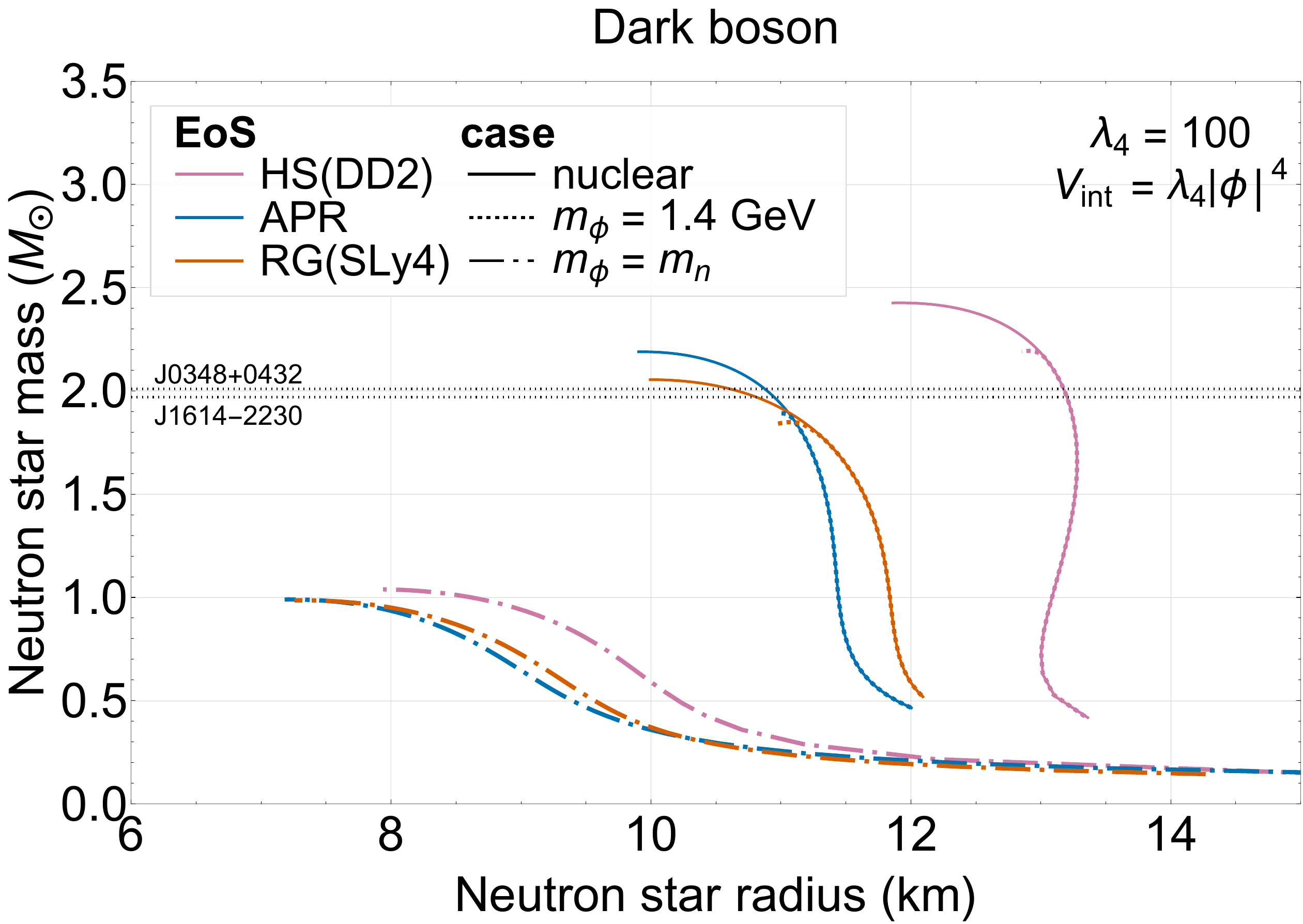}\hfill
    \includegraphics[width=0.48\linewidth]{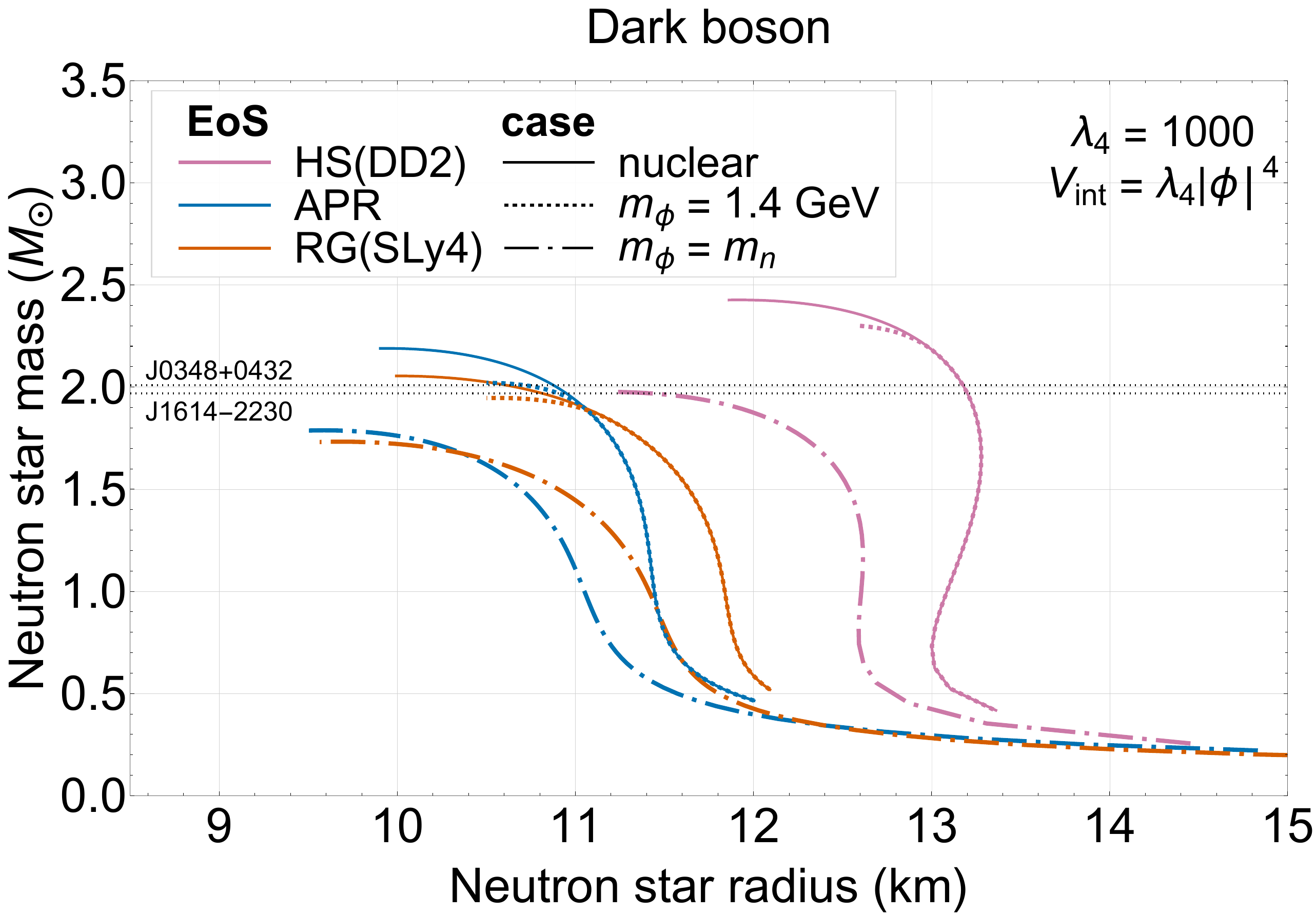}
    \caption{Neutron-star mass--radius curves in presence of a scalar baryon with quartic interaction $V_\text{int} = \lambda_4 |\phi|^4$, where $\lambda_4=100$ (left) and $\lambda_4=1000$ (right). Curves are clipped such that they end at their maximum mass.}
    \label{fig:DarkBoson_quartic}
\end{figure*}

\begin{figure*}[tb]
    \centering
    \includegraphics[width=0.48\linewidth]{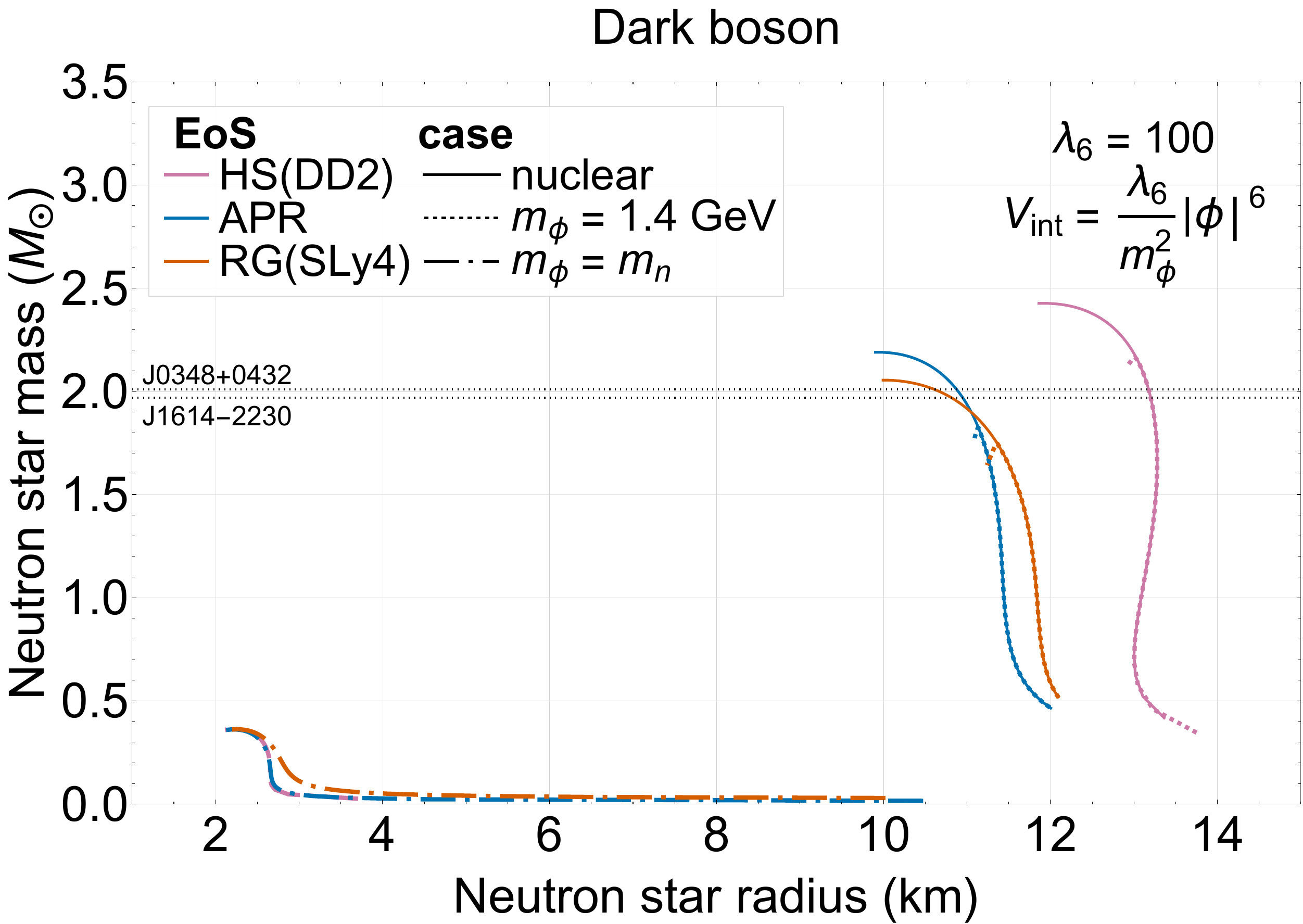}\hfill
    \includegraphics[width=0.48\linewidth]{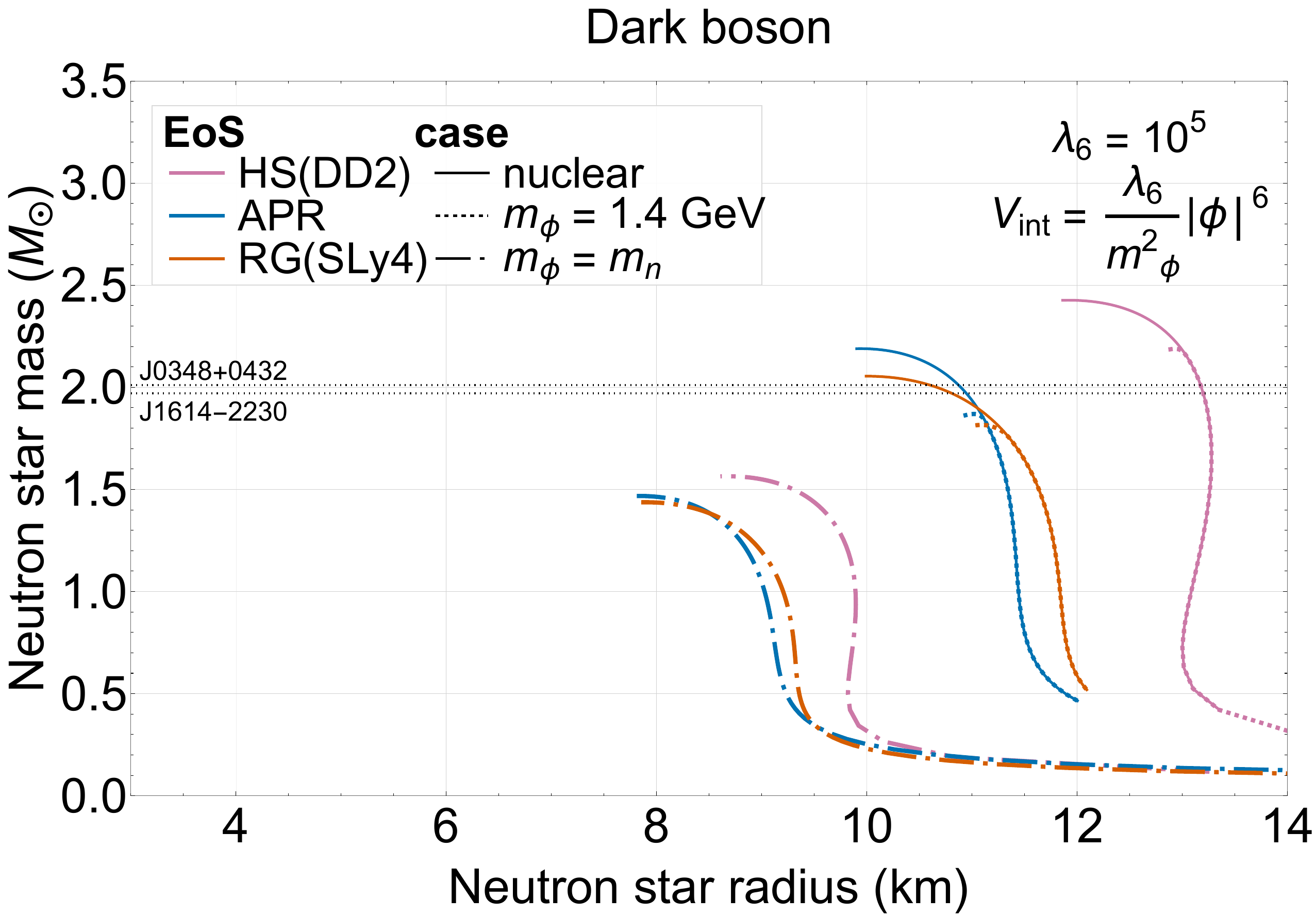}
    \caption{Neutron-star mass--radius curves in presence of a scalar baryon with sextic interaction $V_\text{int} = \frac{\lambda_6}{m_\phi^2} |\phi|^6$, where $\lambda_6=100$ (left) and $\lambda_6=10^5$ (right). Curves are clipped such that they end at their maximum mass.}
    \label{fig:DarkBoson_sextic}
\end{figure*}

Chemical equilibrium identifies $\mu_B = \mu_\phi$, so the neutron star's full EoS in presence of $\phi$ is given by
\begin{align}
    P_\text{tot} &= P_\text{nuc} + P_\phi\,,\\
    \epsilon_\text{tot} &= \epsilon_\text{nuc} + \epsilon_\phi\,,
\end{align}
providing a parametric EoS just like in the dark fermion case. The onset for a $\phi$ population  is again $\mu_B > m_\phi$.

\begin{figure}[tb]
    \centering
    \includegraphics[width=0.95\linewidth]{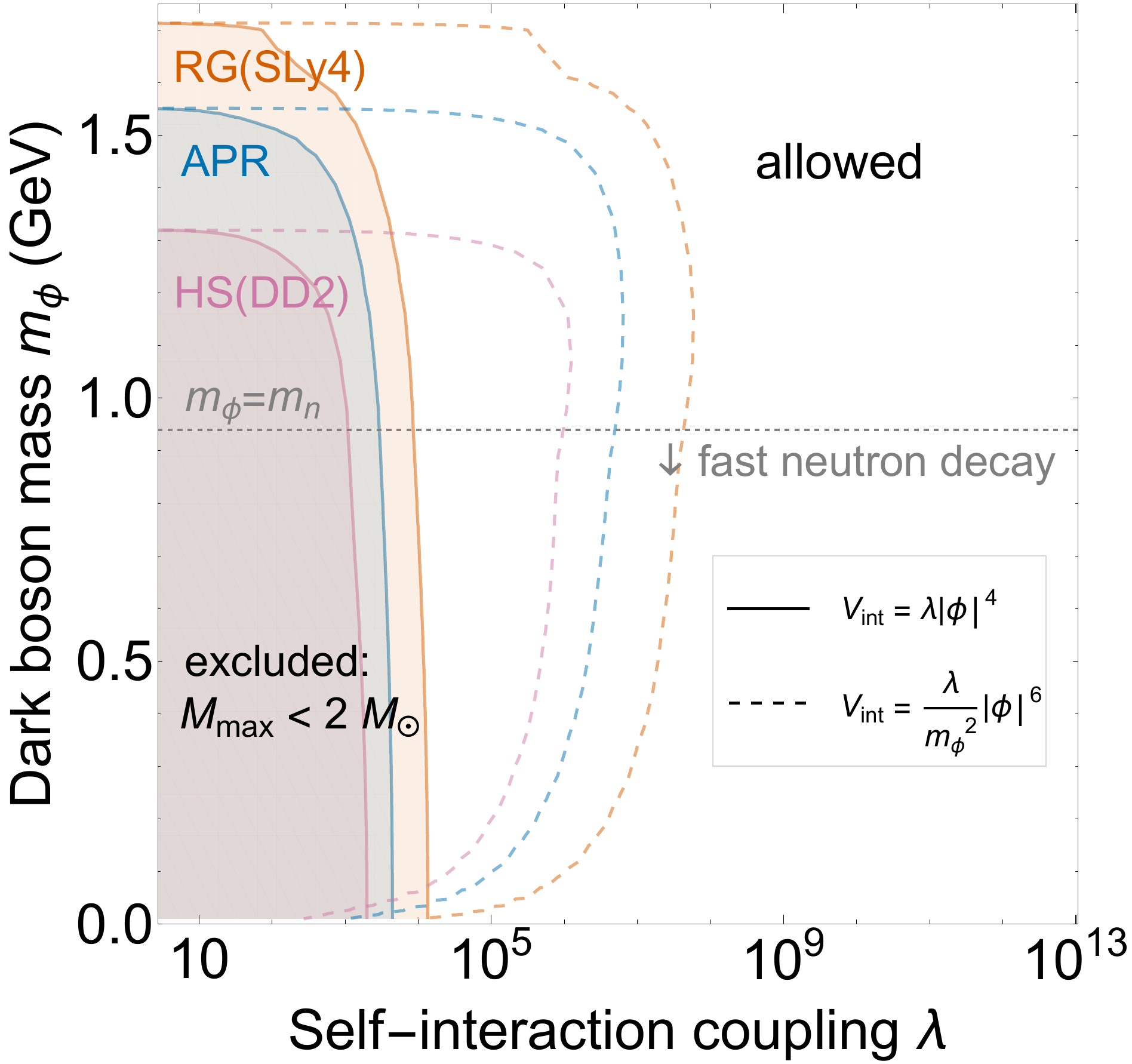}
    \caption{Parameter space excluded by $2 M_\odot$ neutron stars for quartic interactions, $V_\text{int} = \lambda |\phi|^4$, (solid lines) and sextic interactions, $V_\text{int} = \frac{\lambda}{m_\phi^2} |\phi|^6$ (dashed lines). Colors indicate the three chosen nuclear EoS. The region $m_\phi < m_n$, indicated by the gray dotted line, is excluded by fast neutron decays.
    }
    \label{fig:exclusion_monomial}
\end{figure}

The best-motivated case is of course $N=4$, the only renormalizable self-interaction conserving the underlying $U(1)$ symmetry. We show the neutron-star mass--radius relation in the presence of a scalar baryon with quartic interactions in Fig.~\ref{fig:DarkBoson_quartic}. For $m_\phi \leq m_n$, we see that the coupling constant $\lambda_4$ needs to exceed 1000 to push the maximal allowed neutron-star mass above the observed $2 M_\odot$, matching our estimate from above. For $m_\phi > m_n$ the coupling constant can be smaller of course, and ceases to be constrained once $m_\phi > \mu_B^\text{max}$.
$\lambda_4 =100$ yields mass--radius curves qualitatively similar to the dark-fermion case, Fig.~\ref{fig:dark_fermion}, whereas $\lambda_4=1000$ is qualitatively different and much stiffer. In the parameter space that supports $2 M_\odot$ neutron stars, $\phi$ makes up at most $10\%$ of the neutron star, significantly less in the region $m_\phi > m_n$. 
For comparison, we also show the case $N=6$, i.e.~the sextic potential  $V_\text{int} = \frac{\lambda_6}{m^2_\phi}|\phi|^6$, which is not renormalizable but can be viewed as a low-energy effective interaction mediated by heavier particles~\cite{Heeck:2022iky}. The mass--radius curves for this case are shown in Fig.~\ref{fig:DarkBoson_sextic}, which require even larger coupling constants, $\lambda_6 > 10^6$, to realize two-solar-mass neutron stars. We show the exclusion curves in the $m_\phi$--$\lambda$ plane for the quartic and sextic interactions in Fig.~\ref{fig:exclusion_monomial}. In the sextic case, the small-mass regime constrains the effective coupling $\lambda_6/m_\phi^2$ to exceed $\unit[1.1 \times 10^7]{GeV^{-2}}$ (APR), $\unit[2.6 \times 10^6]{GeV^{-2}}$ (HS(DD2)), and $\unit[1.0 \times 10^8]{GeV^{-2}}$ (RG(SLY4)), respectively, which explains the $m_\phi \propto \sqrt{\lambda_6}$ shape of the exclusion curves for $m_\phi\ll m_n$. Notice however that the region $m_\phi < m_n$ is independently excluded by nucleon decays for the large $n\to \phi$ transition rates assumed for chemical equilibrium.
Evading neutron-star constraints for $m_\phi > m_n$ requires either fantastically large repulsive self-interactions or $m_\phi > \mu_B^\text{max}$.
In particular, solving the neutron-lifetime anomaly through the channel $n\to \phi \bar{\nu}$ as in Ref.~\cite{Khatibi:2023fwv} requires additional long-range repulsive interactions on top of the quartic self-interaction term if we stick to perturbative couplings.

\subsection{Polynomial interactions}
\label{sec:Repulsive}

Consider next a repulsive potential with a quartic and a sextic term: $V_\text{int} = \lambda_4 |\phi|^4 + \frac{\lambda_6}{m_\phi^2}|\phi|^6$. The field equation
\begin{equation}
    \mu_\phi^2 - m_\phi^2 = 2\lambda_4 X + 3\frac{\lambda_6}{m_\phi^2}X^2
    \label{eq:polynomial_EOM}
\end{equation}
gives the condensate amplitude 
\begin{equation}
    X = \frac{m_\phi^2}{3\lambda_6} \left[ -\lambda_4 + \sqrt{\lambda_4^2 + \frac{3\lambda_6}{m^2_\phi}(\mu_\phi^2 - m_\phi^2)}\right],
\end{equation}
and the relevant densities are given by
\begin{align}
    P_\phi & = (\mu_\phi^2 - m_\phi^2)X - \lambda_4X^2 - \frac{\lambda_6}{m^2_\phi}X^3\,,\\
    \epsilon_\phi &= 2\mu_\phi^2 X - P_\phi\,.
\end{align}
By comparing the two terms on the right-hand side of Eq.~\eqref{eq:polynomial_EOM}, we see that for $X \ll 2\lambda_4 m_\phi^2/(3\lambda_6)$, the model will resemble the pure quartic monomial case, whereas  $X \gg 2\lambda_4 m_\phi^2/(3\lambda_6)$ resembles the sextic case. 
Defining
\begin{equation}
    \mu_\text{cross} \equiv m_\phi \sqrt{1 + \frac{8\lambda_4^2}{3\lambda_6}}\,,
\end{equation}
the clean criterion is that $|\phi|^4$ dominates when $\mu_B \ll \mu_\text{cross}$, and $|\phi|^6$ dominates when $\mu_B \gg \mu_\text{cross}$. At $\mu_B \sim \mu_\text{cross}$, we have a truly mixed $|\phi|^4 + |\phi|^6$. 
The $\phi$ EoS is quite lengthy, but reduces to $\epsilon_\phi \simeq 2 P_\phi$ for large pressure, just like the sextic case, and $\epsilon_\phi \simeq 2 m_\phi^2\sqrt{P_\phi/\lambda_4}$ for low pressure, just like in the quartic case.
This model does not lead to qualitatively different phenomenology compared to the previous monomial cases, but is a useful comparison for the model below, in which we flip the sign of the quartic interaction term, generating an \textit{attractive} interaction.

\subsection{Attractive interactions}

So far, we have considered only repulsive scalar self-interactions, both because this leads to a bounded-from-below scalar potential and a stable neutron star. To study the effect of \textit{attractive} interactions, we take the interaction potential from the previous section and flip the sign of the quartic term,
\begin{align}
    V_\text{int} = -\lambda_4|\phi|^4 + \frac{\lambda_6}{m_\phi^2}|\phi|^6\,,
    \label{eq:Qball_potential}
\end{align}
with positive $\lambda_4$ and $\lambda_6$. 
To keep the $U(1)$ baryon number symmetry unbroken in vacuum, the scalar potential's global minimum needs to be at $\phi = 0$ and unique, which translates into the strict inequality
\begin{equation}
    \lambda_4^2 < 4\lambda_6\,.
    \label{eq:inequality}
\end{equation}
The repulsive sextic term stabilizes the potential, while the attractive quartic term introduces novel effects; in particular, this classical field theory now allows for stable bound states even without gravitational interactions, in the form of non-topological solitons~\cite{Lee:1991ax} dubbed  Q-balls~\cite{Coleman:1985ki}. For our chosen potential~\eqref{eq:Qball_potential}, those Q-balls have been studied exhaustively in Ref.~\cite{Heeck:2020bau}. In the limit of large, stable Q-balls, the $\phi$ density inside is approximately constant up to the radius, dubbed Q-matter by Coleman~\cite{Coleman:1985ki,Lee:1991ax}.

Following the same steps as before --  just flipping the $\lambda_4$ sign -- we find the condensate amplitude
\begin{equation}
    X = \frac{m^2_\phi}{3\lambda_6}\left[\lambda_4 + \sqrt{\lambda_4^2 + \frac{3\lambda_6}{m^2_\phi}(\mu^2_\phi - m^2_\phi)}\right]
\end{equation}
and the densities
\begin{align}
    P_\phi &= (\mu_\phi^2 - m^2_\phi)X + \lambda_4 X^2 - \frac{\lambda_6}{m^2_\phi}X^3\,,\\
    \epsilon_\phi &= 2\mu_\phi^2 X - P_\phi\,.
\end{align}
In all previous cases, the onset of $\phi$ production required $\mu_\phi > m_\phi$; now, the attractive interaction changes this condition, as the new binding energy can trigger $n\to\phi$ conversion even if $m_\phi > \mu_B$. The pressure $P_\phi$ -- rewritten using the equation of motion -- can now vanish not only in the vacuum phase, $X=0$, but also at the non-zero density 
\begin{equation}
    X_0 = \frac{\lambda_4 m^2_\phi}{2\lambda_6}\,,
\end{equation}
or, equivalently, at the chemical potential
\begin{equation}
    \mu_0 = m_\phi \sqrt{1 - \frac{\lambda_4^2}{4\lambda_6}}\,,
    \label{eq:threshold_chemical_potential}
\end{equation}
which is smaller than $m_\phi$ due to the inequality~\eqref{eq:inequality}.
The scalar baryons will then be produced as soon as the chemical potential exceeds $\mu_0$, so $\mu_B\geq \mu_0$, even though the vacuum mass $m_\phi$ might be larger than $\mu_B$.\footnote{In the Q-ball literature, $\mu_0$ and $\mu_\phi$ are usually denoted by $\omega_0$ and $\omega$, respectively~\cite{Heeck:2020bau}.}

\begin{figure}[tb]
    \centering
    \includegraphics[width=0.95\linewidth]{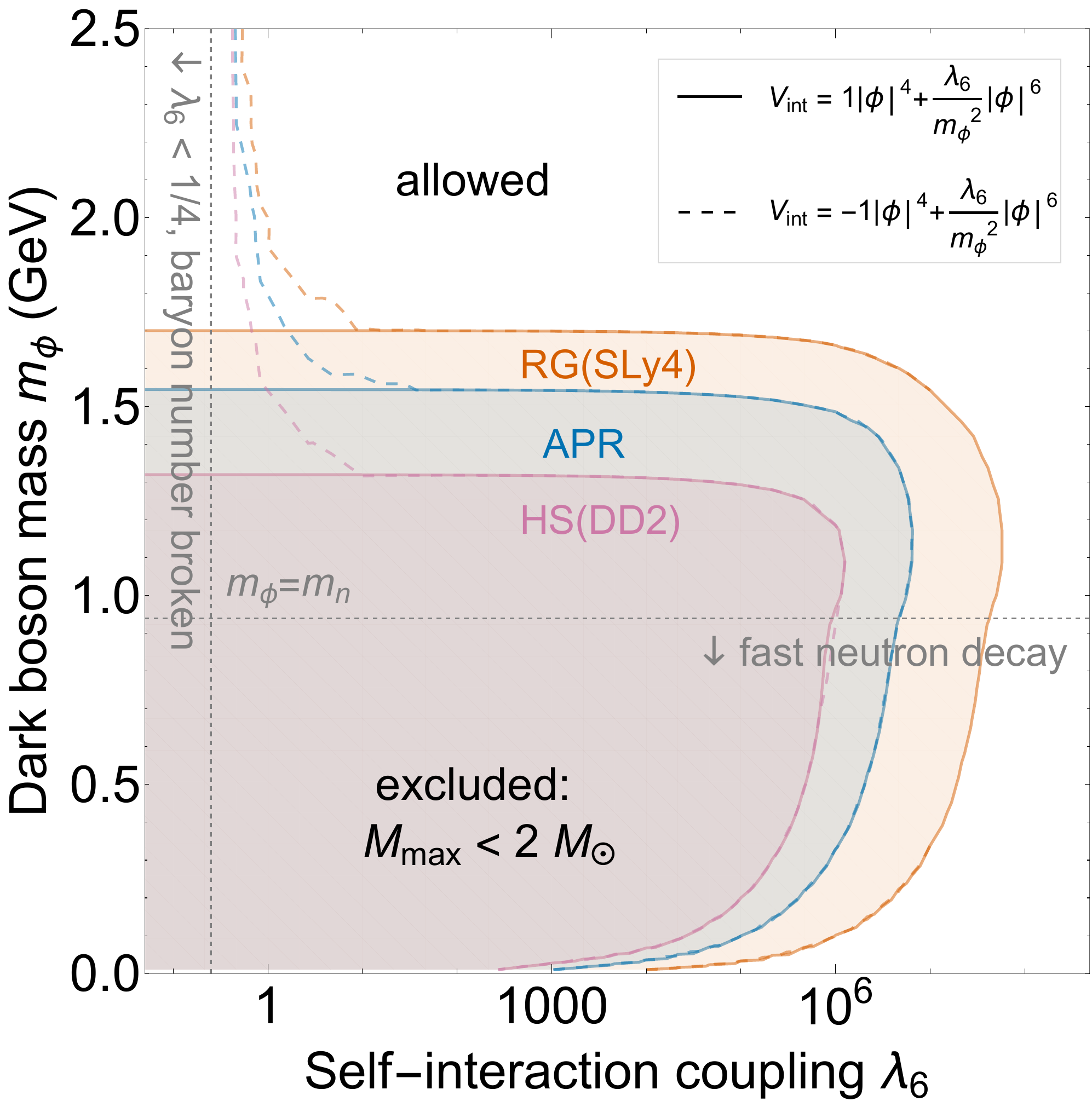}
    \caption{
    Parameter space excluded by $2 M_\odot$ neutron stars for repulsive polynomial interactions (solid lines) and attractive  interactions (dashed lines) with $\lambda_4 = 1$, see text for details. To the left of the vertical gray dotted line the inequality~\eqref{eq:inequality} is violated and baryon number would be spontaneously broken.
    }
    \label{fig:exclusion_polynomial}
\end{figure}

The $\phi$ EoS can be explicitly written in the lengthy form
\begin{align}
\epsilon_\phi &= 2 P_\phi\\
&\ +\frac{m^4\lambda_4}{3\lambda_6}
\left[
1
+2\cosh\!\left(
\frac{1}{3}
\operatorname{arcosh}\!\left(
1+\frac{54\,P_\phi\,\lambda_6^2}{m^4\lambda_4^3}
\right)
\right)
\right]
\nonumber\\[2mm]
&\ -\frac{m^4\lambda_4^3}{36\lambda_6^2}
\left[
1
+2\cosh\!\left(
\frac{1}{3}
\operatorname{arcosh}\!\left(
1+\frac{54\,P_\phi\,\lambda_6^2}{m^4\lambda_4^3}
\right)
\right)
\right]^2,\nonumber
\end{align}
which approximates to $2 P_\phi$ for large pressure -- the familiar result from the sextic case -- and to $\mu_0^2 m_\phi^2 \lambda_4/\lambda_6 + 4 \lambda_6 P_\phi/\lambda_4^2$ for small pressure, unlike any model above and with the hallmark $\epsilon(P=0) = \text{const.}$ for self-bound matter.
We show the limits imposed by $2 M_\odot$ neutron stars in Fig.~\ref{fig:exclusion_polynomial}, together with the purely repulsive polynomial case from the previous section for comparison. For $m_\phi < \mu_B^\text{max}$, there is little difference between an attractive and repulsive quartic interaction, as it is easy enough to produce $\phi$. In this region, $\phi$ makes up less than $15\%$ of the neutron star. However, for $m_\phi > \mu_B^\text{max}$, the repulsive case is not constrained at all, whereas the attractive case is still constrained near $\lambda_4^2\simeq 4\lambda_6$, where $\mu_0\ll m_\phi$ and $\phi$ contributes less than a percent to the neutron star. 
In principle, neutron stars can therefore constrain scalar baryons of arbitrarily large mass in the fine-tuned parameter-space region $\lambda_4^2\simeq 4\lambda_6$; in this phase, the $\phi$ bosons form the same homogeneous Q-matter found in Q-balls, with $\mu_\phi = \mathrm{d}E/\mathrm{d}Q$; a more careful treatment would keep the surface energy that was neglected here.

\section{Conclusions}
\label{sec:conclusions}

Increasingly precise observations of neutron stars allow us to use them as probes of physics beyond the Standard Model. Light new particles in particular better not modify the equation of state too much, lest it fail to allow for the $2 M_\odot$ neutron stars we have detected. This basic constraint has been used in the past on GeV-scale dark \textit{fermions} that carry baryon number and can be produced via neutron transitions if energetically favorable. In this letter, we have adapted this discussion to dark \textit{bosons}, i.e.~scalar baryons. The absence of Fermi pressure makes these light particles even more dangerous to neutron stars, only ameliorated if the repulsive self-interactions have non-perturbatively large couplings. We provided quantitative constraints on repulsive self-interactions for scalar masses up to the maximal chemical potential inside the neutron star, up to $\unit[1.5]{GeV}$ for typical equations of state. In addition, we also considered the case of attractive interactions, with potentials and models familiar from Q-ball solitons; here, the new binding energy allows for production of even heavier scalars, which form a Q-matter condensate.
While we focused on a scalar with baryon number $B=1$, an analogous discussion holds for other cases, e.g.~a scalar $\xi$ with $B=2$ and coupling to two neutrons, $\xi^* n n$~\cite{Heeck:2020nbq,Brugeat:2024rxe}, which can be produced in neutron stars if $m_\xi < 2 \mu_B^\text{max}$, i.e.~up to $\unit[3]{GeV}$.

\section*{Acknowledgments}
This work was supported by the U.S.~Department of Energy under Grant No.~DE-SC0007974.
Data tables of our figures are available as ancillary files of the arXiv version of this article.

\bibliographystyle{utcaps_mod}
\bibliography{BIB.bib}

\end{document}